\documentclass[english,letterpaper,twocolumn,showpacs,pra,aps]{revtex4}
\usepackage{graphicx}
\usepackage{amssymb}

\makeatletter

\large  

\input epsf

\makeatother

\usepackage{babel}
\makeatother
\begin{document}

\title{Casimir energy of a cylindrical shell of elliptical cross section}

\author{Joseph P. Straley$^{1}$, Graham A. White$^{1,2}$ and Eugene B. Kolomeisky$^{3}$}

\affiliation
{$^{1}$Department of Physics and Astronomy, University of Kentucky,
Lexington, Kentucky 40506-0055, USA\\
$^{2}$School of Physics and ARC Centre of Excellence for Particle Physics at the Tera-scale, Monash University, Victoria 3800, Australia \\
$^{3}$Department of Physics, University of Virginia, P. O. Box 400714,
Charlottesville, Virginia 22904-4714,USA}

\begin{abstract}
We calculate the increase in the number of modes (the Kac number) per unit length and the
change in the zero-point energy (the Casimir energy) of the electromagnetic field resulting from
the introduction of a thin perfectly conducting cylindrical shell of elliptical cross-section. Along 
the way we give a novel route to the calculation of these physical quantities.  The Casimir energy is found to be attractive with the circular case corresponding to the energy maximum and the large eccentricity limit being the divergent energy minimum.  As a result, with only Casimir stresses present,  a fixed area shell is unstable with respect to collapse onto itself.  This instability is argued to persist at arbitrary temperature.   
\end{abstract}


\maketitle
\section{Introduction}

The introduction of a conducting surface into a region disturbs the electromagnetic field.  As a general 
measure of the disturbance we can calculate the mode sum
\begin{equation}
\label{modesum}
S(\Omega) = \sum_{\alpha} \left (e^{-\omega_{\alpha}/\Omega} - e^{-\overline{\omega}_{\alpha}/\Omega}\right ) 
\end{equation}
where $\overline \omega_{\alpha}$ are the mode frequencies before introduction of the surface \cite{note1}
and $\omega_{\alpha}$ is the resulting spectrum.  For a general spectrum, the mode sum $S(\Omega)$ is divergent for large values of the parameter $\Omega$ (for example, for a massless scalar field, it diverges as $\Omega^{d-1}$, where $d$ is the dimensionality); however,
for the electromagnetic modes disturbed by a thin perfectly conducting cylindrical shell (of any crosssection), the mode sum per unit length is finite
\cite{BD}.
In this case (the subject of this paper) the quantities of particular interest are the
Kac number per unit length ${\cal K}/L$ (the change in the number of possible modes \cite{kac,BD})
and the Casimir energy per unit length ${\cal C}/L$ (the change in the zero point energy \cite{casimir} formally calculable from the sum
of $\frac {1}{2} (\omega_{\alpha} - \overline\omega_{\alpha})$ \cite{note2}).   These can be extracted from $S$ by 
an expansion in powers of $1/\Omega $:
\begin{equation}
\label{kacnumber}
S(\Omega)/L = {\cal K}/L - \frac {2}{\Omega} {\cal C}/L + ...
\end{equation}
From here the Kac number $\mathcal{K}$ can be recognized as the coefficient $B_{3/2}$ of the heat kernel expansion widely used in the Casimir calculations \cite{heat}.

For the case of a cylinder (of any cross section) the modes have the form
\begin{equation}
\label{cylindermodes}
\omega_{\alpha}(q) = \sqrt{q^2+ w_{\alpha}^2}  ,
\end{equation}
where $q$ is the continuous wave vector of the modes in the axial direction and $w_{\alpha}$ are the cutoff frequencies for the modes.  There are two types of modes, in which either the magnetic field or the electric field is purely transverse \cite{Jackson}.  The cutoff frequencies $w_{\alpha}$ are determined by the 
solutions of the two-dimensional Helmholtz problem for a scalar field representing the component of the electric or magnetic field parallel to the axis of the cylinder, with Dirichlet and Neumann boundary conditions for the two mode types.  The mode sum (\ref{modesum}) is over both types of modes.

The sum over the cutoff frequencies (\ref{modesum}) can be done with the aid of a contour integral representation \cite{contour}.  Let $\psi(w)$ be a function that is zero for each cutoff frequency, and let $\overline \psi(w)$ be the 
corresponding function for the undisturbed space.  Then 
\begin{equation}
\label{contour}
\frac{S(\Omega)}{L} = \frac {1} {2 \pi i} \oint dw 
\int_{- \infty}^{\infty} \frac {dq}{2 \pi} e^{-\sqrt{q^{2}+w^{2}}/\Omega} \frac{d}{dw}
\ln \frac{\psi (w)}{\overline \psi (w)}
\end{equation}
where the first integral is over a contour enclosing the real axis, and $L$ is the (infinite) length of 
the cylinder.  The advantage to this 
representation is that the contour can be deformed to lie along the imaginary $w$ axis, where the 
functions $\psi(w)$ and $\overline \psi(w)$ are slowly varying, instead of highly oscillatory.  The integral over $q$ is
\begin{equation}
\int_{- \infty}^{\infty} e^{-\sqrt{q^2+w^2}/\Omega} dq = 2w K_{1}\left (\frac{w}{\Omega}\right )
\end{equation}
where $K_{n}$ is the modified Bessel function of order $n$. Using this and making the change in variables $w \rightarrow iy$ gives
\begin{eqnarray}
\label{contour2}
\frac{S(\Omega)}{L} &=& \frac {1} {2 \pi^2} \oint dy \ 
 y K_{1}\left (\frac {iy}{\Omega}\right ) \frac {d}{dy}
\ln \frac{\psi (iy)}{\overline \psi (iy)}
\nonumber
\\&=&- \frac {1} {2 \pi^2 \Omega} \int_{-\infty}^{\infty} dy \ 
i y K_{0}\left (\frac {iy}{\Omega}\right ) 
\ln \frac{\psi (iy)}{\overline \psi (iy)}
\end{eqnarray}
In the second representation an integration by parts has been performed, and the contour integral has been converted to an integral over the $y$ axis. 

In the cases to be considered, the function $\ln[\psi (iy)/\overline \psi (iy)]$ 
can be taken to be an even function of $y$.  Then we can replace $i y K_{0}(iy/\Omega)$ by its even part, $(\pi/2) |y| J_{0}(y/\Omega)$ ($J_{n}$ is the Bessel function of order $n$), and reduce the range of integration to the positive $y$ axis, arriving at the representation
\begin{equation}
\label{contour3}
\frac{S(\Omega)}{L} = - \frac {1} {2 \pi \Omega} \int_{0}^{\infty} ydyJ_{0}\left (\frac{y}{\Omega}\right )  \ln \frac{\psi (iy)}{\overline \psi (iy)}
\end{equation}

The quantity $\ln[\psi (iy)/\overline \psi (iy)]$ becomes small for large $y$. 
In general the leading term is of order $y^{-1}$; this determines the Kac number per unit length, as will now be shown.
Define
\begin{equation}
\label{kacdef}
\frac{{\cal K}}{L} = -\frac{1}{2\pi} \lim_{y \rightarrow \infty} y \ln \frac{\psi (iy)}{\overline \psi (iy)}
\end{equation}
and separate Eq.(\ref{contour3}) into two terms:
\begin{eqnarray}
\label{contour4}
\frac{S(\Omega)}{L} &=& \frac{{\cal K}}{L} \int_{0}^{\infty} \frac {dy}{\Omega} J_{0}\left (\frac{y}{\Omega}\right )\nonumber\\
 &-& \frac {1}{\Omega} \int_{0}^{\infty}  dy \ 
 J_{0}\left (\frac{y}{\Omega}\right ) \left (\frac {y} {2 \pi} \ln \frac{\psi (iy)}{\overline \psi (iy)} + \frac{{\cal K}}{L}\right )
\end{eqnarray}
The first integral converges and is equal to unity, independent of the cutoff $\Omega$. The second integral is convergent without the Bessel function, so that we may take the limit $\Omega \rightarrow \infty$ in the integrand.  Then according to Eq.(\ref{kacnumber}), the Casimir energy per unit length is
\begin{equation}
\label{casmir}
\frac{{\cal C}}{L} =  \frac {1}{2} \int_{0}^{\infty} dy 
  \left (\frac {y} {2 \pi} \ln \frac{\psi (iy)}{\overline \psi (iy)} + \frac{{\cal K}}{L} \right)  
\end{equation}

\section{Circular Cylinder}

The Kac number for an arbitrarily shaped conductive shell has been given previously \cite{BD}.  The Casimir energy of the circular cylindrical shell has been also determined \cite{deraad}.  Here we will give a new route to the known result in the familiar setting of the circular cylinder, in preparation for the extension to the case of the cylinder of elliptical cross section.

We consider the effect on the electromagnetic spectrum of the introduction of a cylindrical conducting boundary of radius $A$ into another cylinder of radius $Z$ (which will be taken to infinity shortly \cite{note1}).   The cutoff frequencies of the system interior to the larger cylinder are specified by a scalar function $\varphi(\rho,\phi)$, which satisfies the two-dimensional Helmholtz equation.  The variables can be separated, so that $\varphi$ is a combination of angular factors $\exp(i n \phi)$ and Bessel functions of corresponding order.   There are both $TE$ and $TM$ modes, for the regions $\rho < A$ and $A < \rho < Z$, so that there are four mode conditions: $J_{n}(w A)=0$; $J_{n}'(w A)=0$; $H_{n}^{(1)}(w A) J_{n}(w Z) - H_{n}^{(1)}(w Z) J_{n}(w A)= 0$;
and $H_{n}^{(1)}{'}(w A) J_{n}'(w Z) - H_{n}^{(1)}{'}(w Z) J_{n}'(w A) = 0$ ($H_{n}^{(1)}$ is the Hankel function of order $n$).  Then a candidate for $\psi$ is the product for all $n$ of the left-hand sides of these four expressions, and $\overline \psi$ is the product of the factors $J_{n}(w Z) J_{n}'(w Z)$.  However, for $w = iy$, the Bessel functions are increasing or decreasing exponentially, and in the limit of large $Z$ the exponentially small $H_{n}^{(1)}(iZA)$ terms can be dropped so that $ \psi(iy)/\overline \psi (iy) $
reduces to a product of modified Bessel functions
\begin{equation}
\label{cylinder1}
\frac{\psi(iy)}{\overline \psi (iy)} = \prod_{n} 
-(2 A y)^2 I_{n}(A y) K_{n}(A y) I_{n}'(A y) K_{n}'(A y)
\end{equation}
The factor $-(2Ay)^2$ has been introduced so that the factors in the product approach unity for large $y$.  
This product can be written in a simpler form with the aid of the Wronskian relation $r I_{n}'(r)K_{n}(r)- r I_n(r)K_{n}'(r)= 1$:
\begin{equation}
\label{cylinder2}
\frac{\psi(iy)}{\overline \psi (iy)} = \prod_{n} \left (1 - \sigma_{n}^{2}\right )
\end{equation}
where
\begin{equation}
\label{sigmadefinition}
\sigma_{n} = y \frac {d}{dy} I_{n}(A y) K_n(A y)
\end{equation}

There is an alternate route to (\ref{cylinder1}) that sheds some light on its meaning \cite{KSLZ}.  The path integral representation of quantum mechanics turns three-dimensional quantum mechanics at zero temperature into a four-dimensional classical statistical mechanics problem where the ratio $H/T$ that determines the Boltzmann weight at a temperature $T$ is represented by the ratio of the action to Planck's constant.  The introduction of a boundary suppresses $TM$ modes (because the longitudinal electric field must vanish at the boundary) but introduces new $TE$ modes (because the longitudinal magnetic field can be discontinuous across the boundary).  In either case the effect is localized near the boundary, and can be described by solutions of the modified Helmholtz equation $(\triangle-y^{2})u(\textbf{r},y) = 0$ with sources on the boundary.  These solutions have radial parts that are described by the modified Bessel functions, and the amplitudes of the sources are proportional to $I_{n}(yA)$, $K_n(yA)$, $I_{n}'(yA)$, and $K_{n}'(yA)$ -- the quantities that enter into (\ref{cylinder1}). This approach gives a route to the calculation of the Casimir
energy entirely in terms of functions defined on the
imaginary $w$ axis.
  
Combining (\ref{casmir}) and (\ref{cylinder2}) gives
\begin{equation}
\label{casimircylinder}
\frac{{\cal C}}{L} =  \frac {1}{2}
\int_{0}^{\infty}
\left ( \frac{{\cal K}}
{L} + \frac {y} {2 \pi}  \sum_{n=-\infty}^{\infty}
 \ln (1-\sigma_{n}^2) \right ) dy 
\end{equation}
The logarithm can be expanded in a sum of powers of $\sigma_{n}$.   Then  ${\cal C}/L =  \sum T_{m}$, where
\begin{eqnarray}
\label{cylinderexpansion}
T_{1}&=& \frac {1}{4 \pi } \int_{0}^{\infty}   \left (\frac{2\pi {\cal K}}{L} - y \sum_{n=-\infty}^{\infty}
\sigma_{n}^{2}\right ) dy
\nonumber
\\
T_{m}&=& -\frac {1}{4 \pi m} \int_{0}^{\infty} y dy  \sum_{n=-\infty}^{\infty}
\sigma_{n}^{2m}
\end{eqnarray}

\subsection{Klich Expansion}
 
Klich \cite{klich} has pointed out a useful trick for evaluating the sums that appear in these equations.
The Green function for the modified Helmholtz problem can be expanded in modified Bessel functions about an arbitrary origin, and by equating representations we can derive the identity \cite{JDJ}
\begin{equation}
\label{green}
K_{0}\left (yA \rho(\phi-\phi')\right ) = \sum_{n=-\infty}^{\infty} I_{n}(Ay) K_{n}(Ay) e^{in(\phi-\phi')}
\end{equation}
where 
\begin{equation}
\rho(\phi) = \sqrt{2 - 2 \cos\phi}= 2 \left |\sin \frac{\phi}{2}\right |
\end{equation}
Performing the operation $y d/dy$ on both sides of (\ref{green}) gives
\begin{eqnarray}
\label{Gdef}
H(1,2)&\equiv& y \frac {d}{dy} K_{0}\left (2 y A \left |\sin \frac{\phi_{1}-\phi_{2}}{2}\right |\right )
\nonumber
\\
&=& \sum_{n=-\infty}^{\infty} y \frac{d}{dy} \left (I_{n}(yA) K_{n}(yA)\right ) e^{in(\phi_{1}-\phi_{2})}
\nonumber
\\
&=&\sum_{n=-\infty}^{\infty} \sigma_{n} e^{in(\phi_{1}-\phi_{2})}
\end{eqnarray}
By multiplying this expression times itself $2m$ times, identifying the $\phi_{i}$ in a chain of pairs, and integrating over all $\phi_{i}$ we may derive a series of identities \cite{CM}
\begin{eqnarray}
\label{klichidentities}
\sum_{n=-\infty}^{\infty} \sigma_{n}^{2m} &=& \int_{-\pi}^{\pi}\frac{d\phi_{1}}{2\pi} \ldots
\int_{-\pi}^{\pi}\frac{d\phi_{2m}}{2\pi} H(1,2)\nonumber\\
&\times&H(2,3)\ldots H(2m,1)
\end{eqnarray}
Substituting these into (\ref{cylinderexpansion}) gives an extension of the representation for ${\cal C}$ similar to that given by Balian and Duplantier \cite{BD}. 
 
\subsection{Kac number}
 
For large $y$ the $\sigma_{n}$, Eq.(\ref{sigmadefinition}),  are small, and Eq.(\ref{kacdef}) gives
\begin{eqnarray}
\label{cylK}
 \frac {{\cal K}}{ L} &= &- \lim_{y \rightarrow \infty}
 \frac{y}{2\pi} \ln\frac{\psi(iy)}{\overline \psi (iy)}\nonumber\\
&=& - \lim_{y\rightarrow \infty} \frac{y}{2\pi} \sum_n \ln (1-\sigma_n^2)
=   \lim_{y\rightarrow \infty} \frac{y}{2\pi} \sum_n \sigma_n^2\nonumber\\
&=&  \lim_{y\rightarrow \infty} \frac{y}{2\pi} \int_{-\pi}^{\pi} \frac {d \phi_1}{2\pi}\int_{-\pi}^{\pi} \frac {d \phi_2}{2\pi} H(1,2) H(2,1)
\nonumber
\\
&=&  \lim_{y\rightarrow \infty} \frac{y}{\pi^{2}} \int_{0}^{\pi} d \beta \left [ 2yA\sin \beta K_{1}\left (2yA \sin \beta \right )\right ]^2
\end{eqnarray}
For large $y$, this integral is dominated by the contribution
from small $\beta$, so that we may replace $\sin\beta$
by its argument, leading to \cite{integral}
\begin{eqnarray}
\label{cylK2}
 \frac { {\cal K}}{L} 
&=&  \frac {1}{2\pi^{2}A} \lim_{y\rightarrow \infty}\int_{0}^{2\pi yA} d(2yA\phi) (2yA\phi K_{1}(2yA \phi))^2\nonumber\\
&=& \frac{1}{2\pi^{2}A}\int_{0}^{\infty}dx (xK_{1}(x))^{2}=\frac {3}{64A}
\end{eqnarray}
This result was given previously by Balian and Duplantier \cite{BD}.  They also showed that with this choice $T_{1}$ in (\ref{cylinderexpansion}) vanishes.

\subsection{Casimir energy}

We evaluated the expressions for $T_{m}$ (\ref{cylinderexpansion}) by a Monte Carlo integration.   The first step is to change variables to $y = (\ln z)^{4}$, which compresses the range of integration to the interval $z \in [0,1]$ (the choice of the fourth power is somewhat arbitrary, but was found to give better sampling of the function in the Monte Carlo integration).  Then the multiple integral over the angles and $z$ is given by the average of the integrand (in the new variable) evaluated at uniformly and randomly sampled values for $z$ and the $\phi_{i}$.  Averaging over $10^8$ configurations we find
\begin{eqnarray}
\label{results}
T_2 = \frac{-0.00758 \pm 0.00002}{A^{2}}
\nonumber
\\
T_3= \frac{-0.002264 \pm 0.000002}{A^{2}}
\nonumber
\\
T_4= \frac{-0.001080 \pm 0.000001}{A^{2}}
\nonumber
\end{eqnarray}
For large $m$ the integral is almost entirely due to the terms involving $\sigma_{0} = y d/dy (I_{0}(Ay)K_{0}(Ay))$:
\begin{eqnarray}
\label{results2}
T_{20} =
-\frac{1}{8\pi}\int_{0}^{\infty} \sigma_{0}^4 ydy =-\frac{0.00685}{A^{2}}
\nonumber
\\
T_{30} =
-\frac{1}{12\pi}\int_{0}^{\infty} \sigma_{0}^6 ydy = -\frac{0.002258}{A^{2}}
\nonumber
\\
T_{40} =
-\frac{1}{16\pi}\int_{0}^{\infty} \sigma_{0}^8 ydy = -\frac{0.001081}{A^{2}}
\end{eqnarray}
This series ${ T_{m0} }$ is found to fall off as $0.0346 \times (2m)^{-2.5}$ rather accurately, and it is also possible
to evaluate the sum of the series in the form
\begin{equation}
S_0=
\frac{1}{4\pi} \int_{0}^{\infty} \left (\ln(1-\sigma_{0}^{2}) + \sigma_{0}^{2}\right ) dy = -\frac{0.012799}{A^{2}}
\end{equation}
Therefore only the first few $T_{m}$ need to be calculated;  we thus arrive at the evaluation 
\begin{eqnarray}
\frac{{\cal C}}{L} &=& S_0-T_{20}-T_{30}-T_{40}+T_{2}+T_{3}+T_{4}
\nonumber
\\
&=& \frac{- 0.01354 \pm 0.00002}{A^{2}}
\end{eqnarray}
in agreement with the accepted value ${\cal C}/L = -0.013561343/A^{2}$ \cite{deraad}.  The Monte Carlo program is simple in structure and only requires modest computational resources.  The accuracy is limited by the $T_2$ term, which is difficult to evaluate because the integrand has significant contributions from large $y$ when the four angles $\phi_{i}$ are nearly the same.
 
\section{elliptical cylinder}

\subsection{Mode expansion}
Elliptical coordinates are related to the rectangular coordinates $X, Y$ by
\begin{eqnarray}
\label{elliptical}
X = h \cosh \xi \cos \eta
\nonumber
\\
Y = h \sinh \xi \sin \eta .
\end{eqnarray}
The surfaces of constant $\xi$ are confocal ellipses with axes $A = h \cosh \xi$ and $B = h \sinh \xi$. 
The  distance between the foci is $2h$; the eccentricity of the ellipse is $\epsilon = {\text {sech}} \  \xi$.
The case of the circle is recovered in the limit $\xi \rightarrow
\infty$,  $h \cosh\xi = \sqrt{X^{2}+Y^{2}}$ (which implies $h \rightarrow 0$).  

The two-dimensional modified Helmholtz equation can be separated in these variables.  The solutions have the form of products $P(\eta)Q(\xi)$, where $P$ and $Q$ satisfy the Mathieu equations
\begin{equation}
\label{de-eta}
\frac{d^2 P}{d \eta^2} = -(a +h^2 y^2 \cos^2 \eta) P(\eta)
\end{equation}
\begin{equation}
\label{de-xi}
\frac{d^2 Q}{d \xi^2} = (a + h^2 y^2 \cosh^2 \xi) Q(\xi)
\end{equation}
where $a$ is the separation constant.   The functions $P(\eta;a,y)$ play a role similar to that of the trigonometric functions in the case of the cylinder, and become them in the circle limit.  The condition that $P(\eta;a,y)$ be periodic restricts $a$ to a discrete set of values, but unlike the case of the circle the allowed values are not integers and depend on $y$; we will refer to the set of these as $\{a_{n}(y)\}$,
where $n$ is a counting label (the
modes can be ordered
 so that $a_{n+1} > a_n$, and so that $P(\eta;a_{0},y)$ is 
the nodeless function of $\eta$).
For any $y$, the corresponding set of functions $P(\eta;a_{n}(y),y)$ forms a complete set for expression of a general function of $\eta$; they will be normalized so that for any
$a$ and $b$ in $\{a_n\}$,
\begin{equation}
\label{normaliztion}
\int_{-\pi}^{\pi} P(\eta;a,y) P(\eta;b,y) d\eta = \delta_{ab} 
\end{equation}
Then it follows 
from the completeness of the $P(\eta; a,y)$ that
\begin{equation}
\label{complete}
\sum_{a \in \{a_n\}} P(\eta;a,y) P(\eta';a,y) = \delta ( \eta-\eta')
\end{equation}

The function $Q(\xi;a,y)$ plays a role analogous to a modified Bessel function; there are again two solutions: $Q_{1}$, which is regular at small $\xi$ (and increasing); and $Q_{2}$,  which is small at large $\xi$ (and decreasing).  These are related by the Wronskian relationship
\begin{equation}
\label{MathieuWronskian}
Q_{2}(\xi;a,y) \frac {d}{d\xi} Q_{1}(\xi;a,y)-
Q_{1}(\xi;a,y) \frac {d}{d\xi} Q_{2}(\xi;a,y) = 1
\end{equation}
However, there are some important differences from the case of the circular cylinder.  Eq.(\ref{de-xi}) differs from the Bessel equation in that there is no term in the first derivative.  This implies that the Wronskian is constant, and we are choosing the normalization so that it is unity, as stated in (\ref{MathieuWronskian}).  This also means that the product of the two solutions is asymptotically unity at $\xi$ large, and (via Eq.(\ref{de-xi})) at $y$ large as well.   Additionally the Bessel functions depend on radial position $r$ only through the $ry$ combination, and on what happens in the angular sector only through the order $n$, while the Mathieu function $Q(\xi;a,y)$ depends on $y$, $\xi$, and $a$ separately.  We can make the mathematics of the ellipse look more like the mathematics of the circle if we define the restricted functions $I_{n}(y;\xi) = Q_{1}(\xi;a_{n}(y),y)$ and $K_{n}(y;\xi) = Q_{2}(\xi;a_{n}(y),y)$.  For real cutoff frequency $w$ we can similarly define the oscillatory functions $J_n(w;\xi)$ and $H_{n}(w;\xi)$ that correspond to the usual Bessel functions.

The cutoff frequencies for the elliptical cylinder whose shape and size are determined by $h$ and $\xi$ are determined by the values of $w$ for which either
$J_{n}(w;\xi)= 0$ or $d/d\xi (J_{n}(w,\xi)) = 0$  for the modes inside the cylinder and with a similar but more complicated relationship for the modes outside the cylinder.  The quantity $\psi (w)/\overline \psi (w)$ needed for the contour integral relationship (\ref{contour}) for the mode sum is constructed from these elements as in the circular cylinder case.  After the change of variables $w \rightarrow iy$, we arrive at
\begin{eqnarray}
\label{ellipse1}
\frac{\psi (iy)}{\overline \psi (iy)}&=& -4 I_{n}(y;\xi) \frac {d}{d\xi} I_{n}(y;\xi) K_{n}(y;\xi) \frac {d}{d\xi} K_{n}(y;\xi)
\nonumber
\\
&=&1-\left (\frac {d}{d\xi} (I_n(y;\xi)K_{n}(y;\xi)\right )^{2}
\nonumber
\\
&\equiv&1-\sigma_{n}^2 
\end{eqnarray}
where we have used the Wronskian relationship (\ref{MathieuWronskian}) in the second line.     
The cautious reader will have reason to be suspicious about the analytic continuation of $J_{n}(w;\xi)$ to $I_{n}(y;\xi)$.  In the case of the circular cylinder the order $n$ of the Bessel function is an integer and unaffected by the frequency (its argument $w$); here, the $a_{n}$ depend on the frequency, and take on different values on the real and imaginary axes. For complex values of $y$, the $a_{n}(y)$ are surely also complex.  However, the point is that the Mathieu function $Q(\xi,y,a)$ is analytic in $y$ for any choice of $\xi$ and $a$, and then the restricted function $I_{n}(y;\xi)$ is also analytic;  the way that the set $a_{n}$ evolves as $y$ is varied is built into the definition of the restricted functions.  Finally, we will note that the path integral approach \cite{KSLZ} gives an alternate route to the representation (\ref{ellipse1}).

With this foundation, the determination of the Kac number and the Casimir energy for the cylinder of elliptic cross section is similar to that for the circular cylinder.  In particular, Eqs. (\ref{kacdef})-(\ref{casmir}), (\ref{casimircylinder}), and (\ref{cylinderexpansion}) are unaltered, except that the sums are now over the set $\{a_n(y)\}$.

\subsection{Klich expansion}
In elliptic variables the Green function satisfies the partial differential equation
\begin{eqnarray}
\label{PDE}
\left ( \frac {\partial^2}{\partial \xi^2} +\frac {\partial^2}{\partial \eta^2} 
-y^2 h^2 (\sinh^2 \xi + \sin^2 \eta )\right ) G(\xi,\eta;\xi'
 ,\eta')
\nonumber
\\
=  - 2\pi \delta (\xi-\xi') \delta (\eta - \eta')
\end{eqnarray}
We can represent it in two ways: by direct transcription from the polar variables
\begin{equation}
\label{green1}
G(\xi,\eta;\xi',\eta') =  K_0(y \rho) ,
\end{equation}
where
$\rho$
is the distance between the points $(\xi,\eta)$ and $(\xi',\eta')$; or as an
expansion in Mathieu functions
\begin{eqnarray}
\label{green2}
G(\xi,\eta;\xi',\eta') &=& \sum_{a \in \{a_n\}}
g_n(y)
P(\eta;a,y) P(\eta';a,y)\nonumber\\
 &\times& I_n(\xi,y) K_n(\xi',y)
\end{eqnarray}
for $\xi < \xi'$ (and with $\xi$ and $\xi'$ interchanged for $\xi > \xi'$).
The coefficients $g_n(y)$ are determined by the conditions that $G$ and $\partial G/\partial\xi$ be
continuous at $\xi_1 = \xi_2$ (for $\eta_1 \ne \eta_2$), and that the singularity at $(\xi_1,\eta_1) = (\xi_2,\eta_2)$
has the amplitude implied by (\ref{PDE}).
The result of this analysis is that $g_n(y) = 2\pi$ for all $n$ and $y$.
Thus we have the identity
\begin{eqnarray}
\label{identity}
K_0(y\rho) &=& 2 \pi \sum_{a \in \{a_n(y)\}} P(\eta_1;a,y)P(\eta_2;a,y)\nonumber\\
&\times& I_n(\xi_1,y) K_n(\xi_2,y)
\end{eqnarray}

We only need this result for the case $\xi_1 = \xi_2 = \xi$, where $\rho$ reduces to
\begin{eqnarray}
\label{rhodef}
\rho &=& h\sqrt {\cosh^{2}\xi (\cos \eta - \cos \eta')^2 +\sinh^{2} \xi (\sin \eta - \sin \eta')^2}
\nonumber
\\
&=& h \left |\sin \frac {\eta -\eta'}{2} \right | \Gamma \left (\frac {\eta+\eta'}{2}\right )
\end{eqnarray}
where
\begin{equation}
\label{Gamma}
\Gamma(\alpha) =  \sqrt{\sinh^2\xi+\sin^2 \alpha}
\end{equation}

Take the $\xi$ derivative of (\ref{identity}) to get
\begin{eqnarray}
\label{ident2}
\frac {\partial}{\partial \xi} K_0\left (2 y h \left |\sin\frac {\eta_1-\eta_2}{2}\right | \Gamma\left (\frac {\eta_1+\eta_2}{2}\right )\right ) 
\nonumber
\\
= 2 \pi  \sum_{a \in \{a_n(y)\}} P(\eta_1;a,y)P(\eta_2;a,y) \sigma_n
\end{eqnarray}
Thus by defining
\begin{equation}
\label{Gdef2}
H(\eta_1,\eta_2) =  \frac {\partial}{\partial \xi} K_0\left (2 y h \left |\sin\frac{\eta_1-\eta_2}{2}\right | \Gamma \left (\frac{\eta_1+\eta_2}{2}\right ) \right )
\end{equation}
we obtain an identity equivalent to (\ref{klichidentities}).   
In the circular case (large $\xi$),
$\Gamma \approx \sinh \xi$ and
$\sinh \xi \approx \cosh \xi $, so that (\ref{Gdef2}) reduces to (\ref{Gdef}) with $A = h \cosh \xi$.
For the circular case, 
the Klich expansion was a convenience that improved the rate of convergence by
summing over all orders of the Bessel functions.  In the elliptical case it is even
more useful, because it allows us to do
the sums equivalent to (\ref{cylinderexpansion}) without constructing 
the sets $\{a_n(y)\}$ or the Mathieu functions.

\begin{figure}
\includegraphics[width=1.0\columnwidth,keepaspectratio]{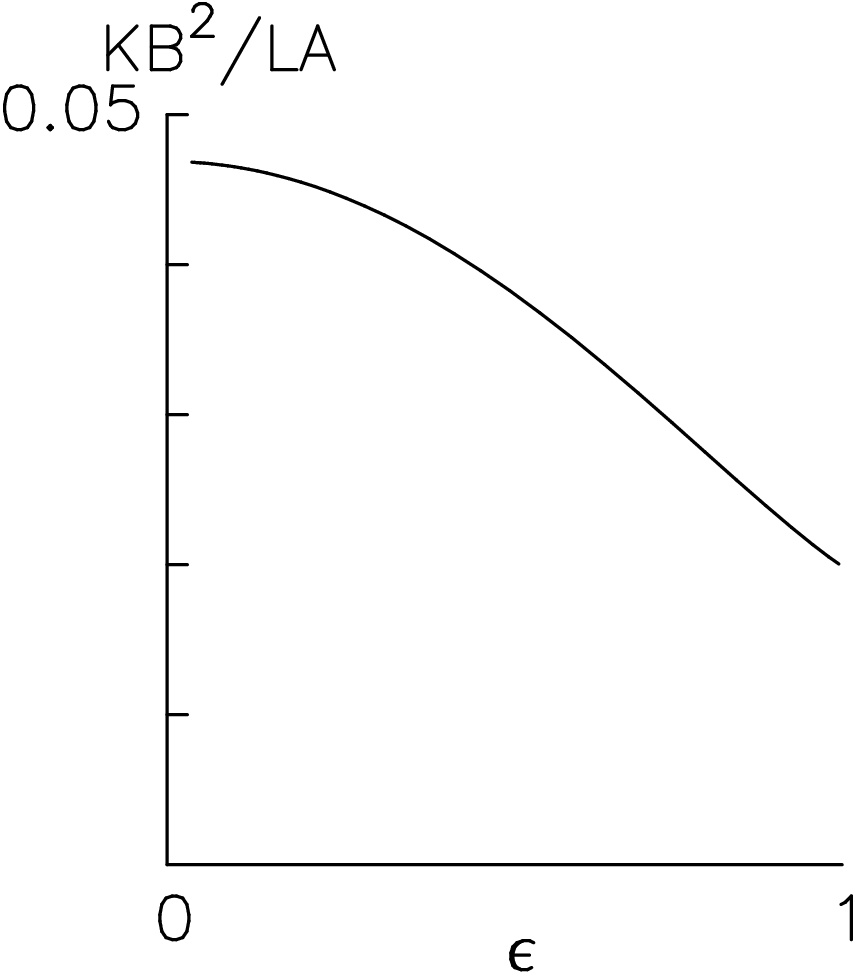} 
\caption{How the Kac number per unit length depends on the shape of the cylinder. }
\end{figure} 

\subsection{Kac number}

Reprising the argument given above, the Kac number per unit length determines the leading
contribution to $\ln[\psi(iy)/\bar \psi(iy)]$ at large $y$.   The analogue to Eq. (\ref{cylK}) is
\begin{eqnarray}
\label{ellipseK}
 \frac { {\cal K}}{L} &=& \lim_{y \rightarrow \infty}
\frac {y}{2\pi}
 \int_{-\pi}^{\pi}
\frac {d\eta_1}{2\pi} \int_{-\pi}^{\pi}\frac {d\eta_2}{2\pi}
H^{2}(\eta_1,\eta_2)
\nonumber
\\
&=&
\lim_{y\rightarrow \infty}\frac {y}{\pi^{3}} \int_{0}^{\pi}
d\alpha \int_{0}^{\pi}d\beta
\left (\frac {d}{d\xi} K_{0}\left (2yh \sin\beta \Gamma(\alpha)\right ) \right )^2\nonumber\\
\end{eqnarray}
where we have changed variables to $\alpha=(\eta_1+\eta_2)/2$ and 
$\beta = (\eta_1 -\eta_2)/2$.  As was previously explained, for large $y$
we can replace $\sin\beta$ by $\beta$; then the integral over $\beta$ is 
the same as was considered in Eqs.(\ref{cylK}) and (\ref{cylK2}).  The result is
\begin{eqnarray}
\label{ellipseK2}
\frac {\cal K}{L} &=& \frac{3}{64h}\int_0^{\pi} \frac{d\alpha}{\pi}
\frac {\sinh^2(\xi) \cosh^2(\xi)}{\Gamma^{5}(\alpha)}
\nonumber
\\
&=& \frac {3}{64} \int_0^{\pi} \frac {d\alpha}{\pi}
\frac {A^2 B^2}
{(A^2 \sin^2 \alpha + B^2 \cos^2 \alpha )^{5/2}}
\end{eqnarray}
which can be evaluated in terms of the complete elliptic integrals ($A > B$ are the
axes of the ellipse).  This agrees from the general expression given
by Balian and Duplantier \cite{BD}.
In the limit of small $B$, this expression is of order $B^{-2}$.  Therefore in Figure 1 we plot ${\cal K}B^2/LA$, which is 
independent of the size of the cylinder and only weakly dependent on its shape.  The horizontal axis is the eccentricity $\epsilon = {\text {sech}} \xi$.

For large eccentricity, the cylindrical shell is approximately a pair of planar strips of width $A$ and area $LA$ separated by a distance $B$.  Then its Kac number must be proportional to the area ($\mathcal{K}\propto LA$) while the Kac number per unit area can only depend on the strip separation $B$.   Since the Kac number itself is dimensionless, the combination of these two arguments predicts a dependence $\mathcal{K}\simeq LA/B^{2}$ thus explaining the divergence of the Kac number at large eccentricity. 

\subsection{Casimir energy}
 
The term $T_1$ in Eq.(\ref{cylinderexpansion}) again vanishes for the case of the cylinder of elliptical cross-section:
substituting (\ref{Gdef2}) and (\ref{klichidentities}) into
(\ref{cylinderexpansion}) and then changing variables from
$\eta_i$ to $\alpha$ and $\beta$ leads to an integral
that differs from the circular case only by a scale factor.
  
The expansion (\ref{cylinderexpansion}) continues to be relevant, as does (\ref{klichidentities}), except
that now $H(\eta_{1},\eta_{2})$ is given by 
(\ref{Gdef2}).  It follows from Eq. (\ref{cylinderexpansion}) that all of the $T_{m}$ are negative, and thus that the Casimir energy is negative.
The numerical evaluation of the $T_{m}$ can be done as above, using almost the same program.  
\begin{figure}
\includegraphics[width=1.0\columnwidth, keepaspectratio]{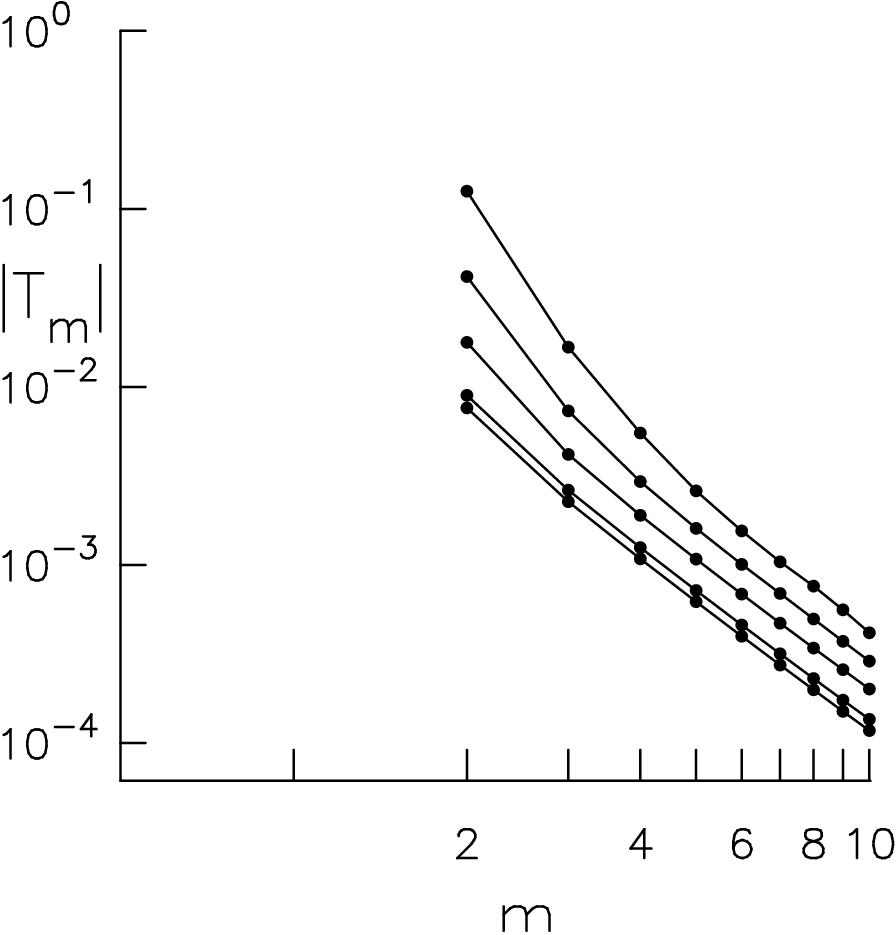} 
\caption{The dependence of $|T_{m}|$ on $m$. 
The curves are drawn for $\epsilon = 0.05, 0.50, 0.80, 0.90, $ and $0.95$, in ascending order.  
The nearly constant slope of these lines indicates that
$-T_{m}$ is decreasing as $m^{-2.5}$.
}
\end{figure} 
Figure 2 shows that it is again true that $-T_{m}$ decreases as $m^{-2.5}$, so that the series
sum converges. 
We evaluated $T_{2}$ through $T_{10}$ by Monte Carlo integration, and applied a truncation correction based on the $m^{-2.5}$ law.  The resulting dependence of ${\cal C}/L$ on the eccentricity $\epsilon$ is shown in Figure 3.
Also shown in Figure 3 is $\sqrt{1-\epsilon^{2}} \times {\cal C}/L$.  The lack of dependence on $\epsilon$ for small $\epsilon$
agrees with the results of Kitson and Romeo \cite{KR}, who discussed the Casimir energy of a cylinder of elliptical cross-section perturbatively.

As was already observed, for large eccentricity the cylindrical shell is approximately a pair of planar strips of width $A$ separated by a distance $B$.  The Casimir attraction between these would give an energy ${\cal C} \simeq - LA/B^3$ \cite{casimir} ($LA$ is the area of a strip); this can be rewritten in terms of the eccentricity as ${\cal C} A^{2}/L \simeq - (1-\epsilon^{2})^{-1.5}$.  The dash-dotted line in Figure 2 shows that this crude argument successfully accounts for the divergence of the Casimir energy at large eccentricity. 
\begin{figure}
\includegraphics[width=1.0\columnwidth, keepaspectratio]{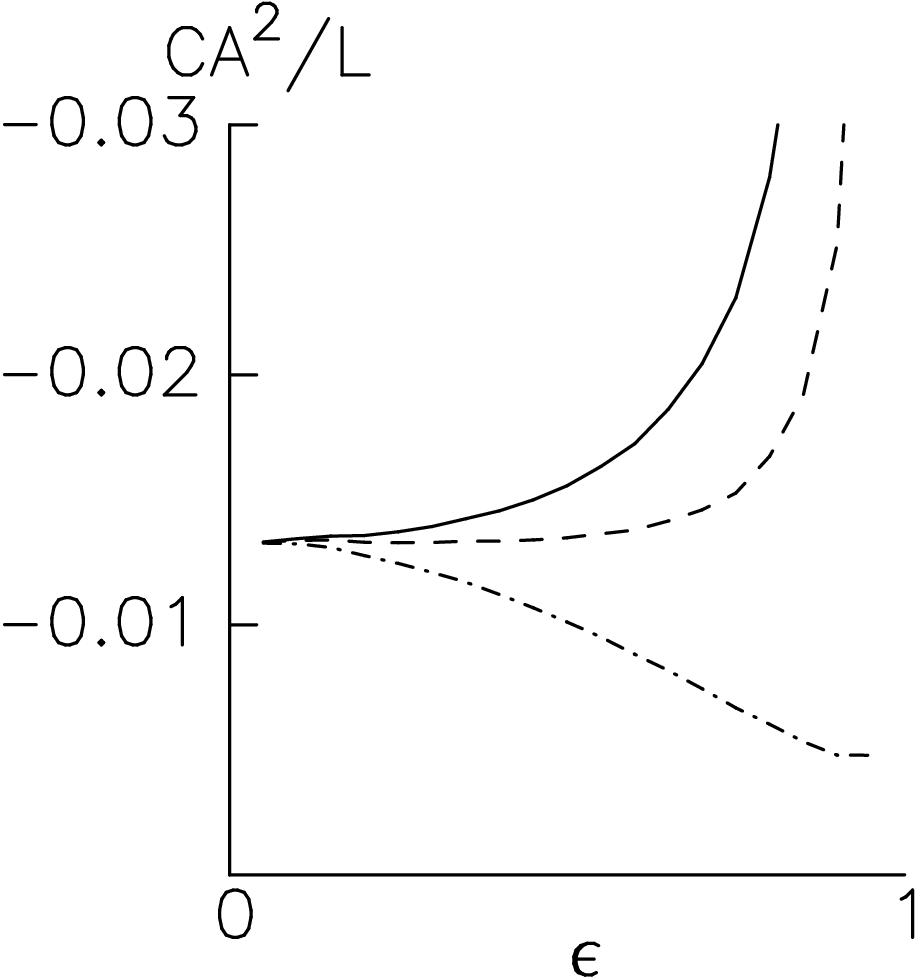} 
\caption{How the Casimir energy per unit length varies with the eccentricity of the cylinder.
The dash-dotted line is $(1-\epsilon^{2})^{1.5} \times |{\cal C}| A^{2}/L$; this removes the
divergence at large eccentricity.
The dashed line is $\sqrt{1-\epsilon^{2}} \times |{\cal C}| A^{2}/L$, which removes the leading order dependence on $\epsilon$ for small $\epsilon$.
}
\end{figure} 

\section{Discussion}

The case of a cylindrical shell with circular cross-section holds a special place in the physics of the Casimir effect.  At zero temperature Casimir forces are known to be attractive for parallel plates \cite{casimir}, repulsive for the sphere \cite{Boyer68, BD} and nearly zero (very weakly attractive) for a long cylindrical shell \cite{BD,deraad}.   Thus the latter is approximately the intermediate case.  The solution of the elliptical cross-section case allows us to follow the evolution of the Casimir attraction with eccentricity $\epsilon$ as the shell cross-section evolves from circular to highly eccentric, resembling the parallel plate geometry.  Therefore it is not surprising that the Casimir energy is found to decrease with eccentricity with the circular case corresponding to the energy maximum.  Naively one might have expected that the Casimir attraction in the $\epsilon\rightarrow 1$ limit would be stronger than its parallel plate counterpart \cite{casimir}. Our results indicate the opposite, however:  even though both attractions have the same order of magnitude, the parallel plate geometry generates stronger attraction.  This illustrates the non-additive character of the Casimir forces.

Since the interactions are attractive with the circular cross-section corresponding to the energy maximum and the large eccentricity $\epsilon=1$ limit being the minimum, with only Casimir stresses present a fixed area shell would be unstable with respect to collapse onto itself.  We argue that this remains the equilibrium state of the system at arbitrary temperature.  Indeed the high-temperature thermodynamics of the system is dominated by the Kac number as the latter determines the form of the free energy \cite{BD}
\begin{equation}
\label{fenergy}
\mathcal{F}\simeq -\mathcal{K}T\ln(Tl)
\end{equation} 
at a temperature $T\gg 1/l$ where $l$ is a length scale approximately corresponding to the largest eigenfrequency of the problem.  For $\mathcal{K}>0$, as is the case in the problem under study, the equilibrium configuration of the shell must have the largest Kac number $\mathcal{K}$.  But we found (Figure 1) that the Kac number diverges in the large eccentricity $\epsilon=1$ limit which corresponds to the collapsed state.  However, as this limit is approached we have $l\simeq B\rightarrow 0$, thus inevitably leaving the range of applicability of the high-temperature expression (\ref{fenergy}) and entering the range of applicability of the zero-temperature theory where the ground state is still collapsed.  This allows us to argue that the equilibrium state of fixed area shell in the presence of Casimir stresses only is collapsed at arbitrary temperature.    

\section{Acknowledgements}

EBK is supported by the US AFOSR Grant No. FA9550-11-1-0297. GAW gratefully acknowledges an Australian Postgraduate Award and the J. L. William Ph. D. scholarship.


\begin{thebibliography}{10}
\bibitem{note1} We imagine the region being studied to be finite by being confined within a large cylinder concentric (or confocal) with the one being introduced, so that the mode frequencies $\overline{\omega}_{\alpha}$ are nonzero.  At an appropriate point in the calculation this constraining boundary will be sent to infinity.
\bibitem{note2}  Throughout this paper we adopt relativistic units $\hbar =1$, $c=1$.
\end{thebibliography}
\end{document}